\documentclass{aastex631}

\usepackage{threeparttable}

\usepackage{mathptmx}
\usepackage{mathrsfs}
\usepackage{xcolor}
\usepackage[mathscr]{euscript}
\usepackage{hyperref}
\usepackage{physics}
\usepackage{simplewick}
\usepackage[bottom]{footmisc}
\usepackage{threeparttable}
\usepackage{appendix}
\usepackage{amssymb}
\usepackage{amsmath}
\usepackage{booktabs} % 导入三线表需要的宏包
\usepackage{threeparttable}
\usepackage{multirow}

\newcommand{\R}[1]{\textcolor{red}{#1}}
\newcommand{\B}[1]{\textcolor{blue}{#1}}
\newcommand{\be}{\begin{equation}}
\newcommand{\ee}{\end{equation}}

\newcommand{\bs}{\begin{split}}
\newcommand{\es}{\end{split}}
\graphicspath{{./}{figure/}}
\usepackage[T1]{fontenc}
\usepackage{subfigure}

\begin{document}

\title{AT2022cmc: a Tidal Disruption Event with Two-component Jet in a Bondi-profile Circumnuclear Medium}

%\title{Constraints on the Circumnuclear Medium Density Profile of AT2022cmc with Multi-band Observations}

\correspondingauthor{Wei-Hua Lei}
\email{leiwh@hust.edu.cn}

\author[0009-0005-9790-1263]{Chang Zhou}
\affiliation{Department of Astronomy, School of Physics, Huazhong University of Science and Technology, Wuhan, 430074, China}

\author[0000-0002-9022-1928]{Zi-Pei Zhu}
\affiliation{Department of Astronomy, School of Physics, Huazhong University of Science and Technology, Wuhan, 430074, China}
\affiliation{Key Laboratory of Space Astronomy and Technology, National Astronomical Observatories, Chinese Academy of Sciences, Beijing, 100101, China}

\author[0000-0003-3440-1526]{Wei-Hua Lei}
\affiliation{Department of Astronomy, School of Physics, Huazhong University of Science and Technology, Wuhan, 430074, China}

\author[0009-0002-7730-3985]{Shao-Yu Fu}
\affiliation{Key Laboratory of Space Astronomy and Technology, National Astronomical Observatories, Chinese Academy of Sciences, Beijing, 100101, China}
\affiliation{University of Chinese Academy of Sciences, Chinese Academy of Sciences, Beijing 100049, China}

\author[0000-0001-5553-4577]{Wei Xie}
\affiliation{Department of Astronomy, School of Physics and Electronic Science, Guizhou Normal University, Guiyang, 550001, China}
\affiliation{Guizhou Provincial Key Laboratory of Radio Astronomy and Data Processing, Guizhou Normal University, Guiyang, 550001, China}

\author[0000-0003-3257-9435]{Dong Xu}
\affiliation{Key Laboratory of Space Astronomy and Technology, National Astronomical Observatories, Chinese Academy of Sciences, Beijing, 100101, China}

\begin{abstract}
A supermassive black hole can launch a relativistic jet when it violently disrupts a star that passes too close. Such jetted tidal disruption events (TDEs) are rare and unique tools to investigate quiescent supermassive black holes, jet physics, and circumnuclear environment at high redshift. The newly discovered TDE AT2022cmc ($z\sim 1.193$) providing rich multi-band (X-ray, UV, optical, sub-millimeter, and radio) data, has been interpreted as the fourth on-axis jetted TDE. In this work, we constrain the circumnuclear medium (CNM) density profile with both closure relation (CR) test and detailed forward shock model fit with Markov chain Monte Carlo (MCMC) approach to the multi-band (optical, sub-millimeter, and radio) data of AT2022cmc.
%\textbf{These rare events--known as jetted tidal disruption events (TDEs)--offer a unique tool} to investigate quiescent supermassive black \textbf{holes}
%in order to constrain the circumnuclear medium (CNM) density profile of AT2022cmc, we employ first the closure relation (CR) test and then MCMC fit forward shock code to the optical and radio data.
%to test the circumnuclear medium (CNM) density profile, and then perform multi-band fit to the radio and optical data with our forward shock code. 
%We find the jet expands into a circumnuclear medium (CNM) of density profile $n\propto R^{-k}$ with $k=1.8\pm0.1$, consisting with a Bondi-like profile. 
We find that the CNM density profile of AT2022cmc is $n\propto R^{-k}$ with $k \sim 1.68$, implying a Bondi accretion in history. Furthermore, our model fit result suggests a two-component jet in AT2022cmc, indicating a similar jet physics to well-studied jetted TDE Sw J1644+57. 
\end{abstract}

\keywords{\href{http://astrothesaurus.org/uat/1696}{Tidal disruption (1696)}}

\section{Introduction} \label{sec:intro}

Tidal disruption events (TDEs) are transient events when a star comes too close to a supermassive black hole (SMBH) to be torn apart by the black hole \citep{Rees1988,Phinney1989}. TDEs have been observed in various wavelengths (radio, UV/optical, X-rays, even $\gamma$-ray in some jetted TDEs), providing a new probe to study SMBH of quiescent galaxies.

The detections of Sw J1644+57 \citep{Bloom2011,Burrows2011,Levan2011,Zauderer2011}, Sw J2058+05 \citep{Cenko2012} and Sw J1112-82 \citep{Brown2015} suggested that at least some TDEs can launch a relativistic jet toward Earth. Recent study on IGR J12580+0134 revealed the first TDE with an off-axis relativistic jet \citep{Lei2016,Yuan2016}. The on-axis jetted TDEs are rare, but are radio-loud \citep{Alexander2020}, and have special use in unveiling cosmological quiescent SMBHs, jet physics, and density profile of pre-existing CNM at $z>1$.
%and are ideal test beds to understand the radiative mechanisms operating in super-Eddington jets

The CNM density profile\,($n\propto R^{-k}$) provides key diagnostics of the accretion history of the SMBH. Currently, there are three special types of CNM density profiles: 1) Sgr $\rm A^{*}$-like\,($k=1$), namely, the CNM density profile consistent with Sgr $\rm A^{*}$ whose $k=1$ \citep{Baganoff03}, AT2019dsg is one of the TDEs satisfying this distribution \citep{Stein21}; 2) Bondi-like\,($k=3/2$), consistent with Bondi accretion whose $k=3/2$, Sw J1644+57 is one of the TDEs satisfying this distribution \citep{Eftekhari18}; 3)\,ASASSN-14li-like ($k=5/2$), there are several TDEs satisfying this distribution, such as ASASSN-14li \citep{Alexander16}, CNSSJ0019 \citep{Anderson20} and Arp 299 \citep{Mattila18}. Such a steep profile is not expected for spherically symmetric accretion, but appears in some models of super-Eddington accretion flows \citep[ZEBRAs]{Coughlin2014}. The accretion history of a dormant SMBH especially at $z>1$ is of great interest.

Recently, a very luminous TDE AT2022cmc has been discovered in optical at $z=1.193$. Its unusual X-ray behavior (peak luminosity of $\geq 3\times 10^{47} \rm erg \ s^{-1}$, short variability timescales of $\sim 1000$ s, and duration of $>30$ days) and long-lived radio/mm data (consistent with synchrotron radiation) indicated that AT2022cmc is an on-axis jetted TDE \citep{Andreoni2022,Pasham2023}.
%The multi-band observations (X-ray, optical, sub-millimeter and radio) supports the interpretation of AT2022cmc as a jetted TDE. 
The rich multi-band observational data offer a good opportunity to explore the properties of SMBH and the CNM density distribution.
For a jetted TDE, the multi-band afterglow emission is produced by the external shock when the jet interacts with the CNM. Therefore, we can use the afterglow observations of TDEs to infer the CNM density at parsec or even sub-parsec scales \citep{Alexander16}. As the jet expands, the afterglow emissions at different epochs can be used to construct the CNM density profile. Such constraints are extremely valuable, as these scales are not directly resolvable at any wavelength with current facilities at the distance of most TDE hosts.
%otherwise unresolvable

Based on equipartition analysis \citep{Chevalier1998,Barniol2013} and afterglow modeling, \citet{Matsumoto2023} roughly estimated the density profile of AT2022cmc as $k\simeq 1.5-2$, which is similar to Sw J1644+57 and generally consistent with a Bondi-like accretion. However, a good constraint on density profile index $k$ is expected for a better understanding of the accretion history of SMBH. On the other hand, the detailed studies on the X-ray and radio observations showed a two-component jet in Sw J1644+57 \citep{Wang2014,LiuDangbo2015,Mimica2015}. The investigation of the jet structure of AT2022cmc is also desired.

In this paper, instead of the equipartition method, we employ a forward shock (FS) model similar to that used in gamma-ray burst (GRB) afterglows. Different from \citet{Matsumoto2023} and \citet{Yao2023}, we first present the closure relations (CRs) of jet-CNM interaction with an arbitrary CNM density profile \citep{Eerten2009,Fraija2021}, and analyze the radio light curve and spectrum with the closure relations. In this way, we can get a rough estimate of the CNM density profile index $k$. Then, we perform a detailed forward shock model fit to the multi-band (optical, sub-millimeter, and radio) data of AT2022cmc with Markov chain Monte Carlo (MCMC) method and obtain a better constraint on the profile index $k$. Finally, we adopt a two-component jet model as in \citet{Wang2014} but modified with an arbitrary CNM density profile to fit the data.

Like Sw J1644+57, the early rapidly-declining X-rays are also assumed to arise from the internal emission of the jet \citep{Matsumoto2023}. Following \citet{Andreoni2022}, the blue optical plateau phase should be a thermal emission unrelated to the jet-CNM interaction. The late-time r-band ($\geq 5$ days) data is likely a thermal origin similar to other optically-selected TDEs. Therefore, for the optical data, we only use the early r-band ($< 5$ days) data in our multi-band afterglow fit, as did in \citet{Matsumoto2023}. 

The paper is organized as follows: The analysis with closure relations is presented in Section\,\ref{sec:CR}. We constrain the CNM density profile index $k$ of AT2022cmc by using the MCMC fit to the radio, sub-millimeter, and optical data in Section\,\ref{sec:model}. In Section\,\ref{sec:disussion}, we discuss the results, e.g., the accretion history, two-component jet origin, and detectability of AT2022cmc-like events. The main conclusions are summarised in Section\,\ref{sec:summary}. A standard cosmology model is adopted with $H_{0}=67.3\ \rm{km\cdot s^{-1} Mpc^{-1}}$, $\Omega_{M}$=0.315, $\Omega_{\Lambda}$=0.685 \citep{Planck+2014}.

\section{Closure Relation Analysis} \label{sec:CR}

The afterglow (radio, sub-millimeter, and early optical) of AT2022cmc can be interpreted with synchrotron emission from the forward shock (FS) of a relativistic jet propagating into the CNM \citep{Matsumoto2023}. The synchrotron flux can be described by a series of power-law segments $F_\nu \propto t^{-\alpha} \nu^{-\beta}$ \citep{Sari1998,Gao+2013,Zhang2018}. In such a model, the type of CNM can be tested with the closure relations (CRs, relations between the temporal indices $\alpha$ and spectral indices $\beta$), as did in gamma-ray bursts (GRBs) \citep{Gao+2013}. \citet{Gao+2013} only presented the CRs for constant-density ($k=0$) and wind type ($k=2$). In this work, to study the CNM type of AT2022cmc, we need to obtain the CRs for an arbitrary density profile as described in several previous works \citep{Eerten2009, Fraija2021} .

To do this, we consider a relativistic thin shell with energy $E_{\rm K,iso}$, initial Lorentz factor $\Gamma_0$, and opening angle $\theta_{\rm j}$ expanding into the pre-existing CNM with density $n$ \citep{Rees1992,Meszaros1997,Sari1998,Zhang2018,Huang2021}, 
\begin{equation}
    n(R)=n_{18} \left( \frac{R}{10^{18} \rm cm} \right)^{-k}=A R^{-k},
\label{eq:CNM_profile}
\end{equation}
where $n_{18}$ is the CNM density at distance $R=10^{18} \rm cm$. We adopt an analytical description of the main properties of the evolution and emission of the FS from the jet-CNM interaction. 
%In this work, to constrain the type of CNM profile based on the analysis with CRs, we should first obtain the CRs for arbitrary CNM distribution, 

%First, we constrain the type of CNM profile based on the analysis with closure relations. To obtain the closure relations for different CNM distributions, we provide an analytical description of the main properties of the evolution and emission of FS. Generally, the evolution includes four phases. 

Generally, the evolution of the jet includes four phases. The first is a coasting phase, in which we have $\Gamma(t)\simeq \Gamma_0$. In the second phase, the shell starts to decelerate 
%at the deceleration time 
%\begin{equation}
%t_{\rm dec}= \left( \frac{3E_{\rm K,iso}}{16\pi n_1 m_{\rm p} \Gamma_0^8 c^5}  \right)^{1/3}    
%\end{equation}
%After $t_{\rm dec}$, 
when the mass $m$ of the CNM swept by the FS is about $1/\Gamma_0$ of the rest mass in the ejecta $M_{\rm ej}$. The shell then approaches the \citet{BM1976} self-similar evolution. The \citet{BM1976} solution for arbitrary $k$ density profile is given by, 
\begin{equation}
\Gamma(t)\simeq \left(\frac{(17-4k)E_{\rm K,iso}}{4^{5-k} (4-k)^{3-k} \pi A m_{\rm p} c^{5-k} t^{3-k} }  \right)^{\frac{1}{2(4-k)}}, \ \ \  R(t)\simeq \left(\frac{(17-4k)(4-k) E_{\rm K,iso} t}{4 \pi A m_{\rm p} c}  \right)^{\frac{1}{(4-k)}} ,
\label{eq:BM}
\end{equation}

\noindent where $m_{\rm p}$ is proton mass. Later, as the ejecta is decelerated to the post-jet-break phase 
%at the time 
%\begin{equation}
%t_{\rm j}\simeq 0.6 {\rm day} \left(\frac{\theta_{\rm j}}{0.1 {\rm rad}}  \right)^{8/3}  \left(\frac{E_{\rm K,iso}}{10^{53} {\rm erg}} \right)^{1/3} n_1^{-1/3},
%\end{equation}
when the $1/\Gamma$ cone becomes larger than $\theta_{\rm j}$. Finally, the blastwave enters the Newtonian phase when it has swept up the CNM with the total rest mass energy comparable to the energy of the ejecta. The dynamics is described by the well-known Sedov-Taylor solution. 

During the dynamical evolution of the FS, electrons are believed to be accelerated at the shock front to a power-law distribution $N(\gamma_{\rm e}) \propto \gamma_{\rm e}^{-p}$. Assuming a fraction $\epsilon_{\rm e}$ of the shock energy $e_2=4\Gamma^2 n m_{\rm p} c^2$ is distributed into electrons, this defines the minimum injected electron Lorentz factor \citep{Zhang2018},
\begin{equation}
\gamma_{\rm m}=\frac{p-2}{p-1} \epsilon_{\rm e} (\Gamma-1)\frac{m_{\rm p}}{m_{\rm e}}    
\end{equation}
where $m_{\rm e}$ is electron mass. We also assume that a fraction $\epsilon_{\rm B}$ of the shock energy is in the magnetic field generated behind the shock. This gives the comoving magnetic field \citep{Sari1998}
\begin{equation}
B=(32\pi m_{\rm p} \epsilon_{\rm B} n)^{1/2} c.   
\end{equation}
The synchrotron power and characteristic frequency from an electron with Lorentz factor $\gamma_{\rm e}$ are given by \citep{Rybicki1979}
\begin{equation}
P(\gamma_{\rm e})\simeq \frac{4}{3} \sigma_{\rm T} c \Gamma^2 \gamma_{\rm e}^2 \frac{B^2}{8\pi},   
\end{equation}
\begin{equation}
\nu(\gamma_{\rm e}) \simeq \Gamma \gamma_{\rm e}^2 \frac{q_{\rm e} B}{2\pi m_{\rm e}c (1+z)},
\end{equation}
where $P(\gamma_{\rm e})$ is expressed in the source frame, $\nu$ is measured in the observer frame, $\sigma_{\rm T}$ is the Thomson cross-section, $q_{\rm e}$ is electron charge. The peak of the  spectra power occurs at $\nu(\gamma_{\rm e})$, and \citep{Sari1998,Gao+2013,Zhang2018}
\begin{equation}
P_{\nu,{\rm max}} \simeq \frac{P(\gamma_{\rm e})}{\nu(\gamma_{\rm e}) (1+z)}=\frac{m_{\rm e}c^2 \sigma_{\rm T}}{3q_{\rm e}} \Gamma B.   
\end{equation}
By equating the lifetime of electron to the time $t$ (in observer frame), one can define a critical electron Lorentz factor $\gamma_{\rm c}$ \citep{Rybicki1979,Sari1998}
\begin{equation}
\gamma_{\rm c} = \frac{6\pi m_{\rm e}c}{\Gamma \sigma_{\rm T}B^2 t/(1+z)},
\end{equation}
the electron distribution shape should be modified for $\gamma_{\rm e} >\gamma_{\rm c}$ when cooling due to synchrotron radiation becomes significant. Accounting for the radiative cooling and the continuous injection of new accelerated electrons at the shock front, one expects a broken power-law energy spectrum of them, which leads to a multi-segment broken power-law radiation spectrum separated by three characteristic frequencies at any epoch \citep{Gao+2013,Zhang2018}. The first two characteristic frequencies $\nu_{\rm m}$ and $\nu_{\rm c}$ in the synchrotron spectrum are defined by the two electron Lorentz factors $\gamma_{\rm \,m}$ and $\gamma_{\rm \, c}$, respectively. The third characteristic frequency is the self-absorption frequency $\nu_{\rm a}$, below which the synchrotron photons are self-absorbed. It can be calculated in two different ways. The first one is the optical depth method by the condition that the photon optical depth for self-absorption is unity ($\alpha_\nu(\nu_a)\Delta\sim 1$, where $\Delta$ is the characteristic width of the emission region) \citep{Rybicki1979}. Another way is the blackbody method by equating the synchrotron flux and the flux of a blackbody ($I_{\nu}^{bb}(\nu_a)=I_{\nu}^{syn}(\nu_a)=2kT\frac{\nu_a^2}{c^2}$) \citep{Gao+2013,Zhang2018}. It can be proved that the two methods are equivalent to each other \citep{Shen2009}. 

The maximum flux density is $F_{\nu,{\rm max}}=(1+z) N_{\rm e} P_{\nu,{\rm max}}/4\pi D^2$ \citep{Sari1998} , where $N_{\rm e}=\int 4\pi R^2 n dR$ is the total number of electrons in shocked CNM and $D$ is the distance of the source.

In this work, following \citet{Matsumoto2023}, the early r-band data is attributed to the synchrotron emissions of the FS. The peak of the early optical phase ($\le 1$ day) can be reasonably explained as the onset of the deceleration phase. We thus focus on the second phase, i.e., the self-similar deceleration phase. The spectra is likely in $\nu_{\rm a}<\nu_{\rm m}<\nu_{\rm c}$ regime. 
%Note that we just employ the CR analysis on the radio observation data at the $\nu_{\rm a}<\nu_{\rm m}<\nu_{\rm c}$ regime. 
%Meanwhile, we roughly estimate the characteristic lorentz factors and find that the spectra of AT2022cmc during $\thicksim8\,-\thicksim60$ days are correspond to $\nu_{\rm a}<\nu_{\rm m}<\nu_{\rm c}$ regime, and most of its data we got is at this phase. So we just employed the CR analysis on the data during $\thicksim8\,-\thicksim60$ days. 
From Equation (\ref{eq:BM}), $\Gamma \propto t^{-\frac{(3-k)}{2(4-k)}}$ and $R\propto t^{\frac{1}{4-k}}$, one has the scalings for the FS spectra parameters in this phase as (for $\nu_{\rm a}<\nu_{\rm m}<\nu_{\rm c}$)
\begin{equation}
\nu_{\rm a} \propto t^{-\frac{3k}{5(4-k)}}, \ \ \nu_{\rm m}\propto t^{-3/2},\ \ \nu_{\rm c} \propto t^{- \frac{4-3k}{2(4-k)} }, \ \  F_{\nu, {\rm max} } \propto t^{-\frac{k}{2(4-k)}} .  
\end{equation}
We can then obtain,
\begin{eqnarray}
F_{\nu} =F_{\nu, {\rm max}} \times \left\lbrace
\begin{tabular}{l}
$\left(\frac{\nu_{\rm a}}{\nu_{\rm m}} \right)^{\frac{1}{3}}  \left(\frac{\nu}{\nu_{\rm a}} \right)^2  \propto t^{\frac{2}{4-k}} \nu^2  , \ \ \ \nu<\nu_{\rm a}<\nu_{\rm m}<\nu_{\rm c}$  \\
$\left( \frac{\nu}{\nu_{\rm m}}  \right)^{\frac{1}{3}} \propto t^{\frac{2-k}{4-k}} \nu^{\frac{1}{3}} , \ \ \ \nu_{\rm a}<\nu<\nu_{\rm m}<\nu_{\rm c}$ \\
$\left( \frac{\nu}{\nu_{\rm m}}  \right)^{-\frac{p-1}{2}} \propto t^{-\frac{12p-12+5k-3kp}{16-4k}} \nu^{-\frac{p-1}{2}}  , \ \ \ \nu_{\rm a}<\nu_{\rm m}<\nu<\nu_{\rm c}$  \\
$\left(\frac{\nu_{\rm c}}{\nu_{\rm m}} \right)^{-\frac{p-1}{2}}  \left(\frac{\nu}{\nu_{\rm c}} \right)^{-\frac{p}{2}} \propto t^{-\frac{3p-2}{4}} \nu^{-\frac{p}{2}} , \ \ \ \nu_{\rm a}<\nu_{\rm m}<\nu_{\rm c}<\nu$
\end{tabular} 
\right.
\end{eqnarray}
which are consistent with the results of \citet[see Tables 1 and 2 therein]{Eerten2009}, but the latter did not include the self-absorption ($\nu_{\rm a}$) \footnote{In the $\nu_a<\nu_m<\nu_c$ regime, the expression for $\nu_{\rm a}$ obtained in our work is  \begin{small}
\begin{equation}
\begin{aligned}
\nu_a= 2.3\times10^{11} {\rm Hz} \frac{g(p)}{g(2.3)} \frac{f(k)}{f(2)}   (1+z)^{\frac{4(5-2k)}{5(k-4)}} n_{18}^{\frac{12}{20-5k}} E_{\rm K,iso,52}^{\frac{4(k-1)}{5(k-4)}}\epsilon_{e,-1}^{-1}\epsilon_{B,-2}^{1/5} t_5^{\frac{3k}{5(k-4)}} \nonumber
\end{aligned}
\end{equation}
\end{small}
where $g(p)=\frac{(p-1)^{8/5}}{p-2}(\frac{p+2}{3p+2})^{3/5}$, \, $f(k)=2^{\frac{6+199k}{20-5k}}\times5^{\frac{201k}{20-5k}}\times(2.998\times10^{10})^{\frac{5-k}{5(k-4)}}\times(1.90605\times10^{75})^{\frac{1}{20-5k}} \times e^{\frac{118.263k-88.6973}{k-4}}\times (17-4k)^{\frac{4(k-1)}{5(k-4)}}\times(4-k)^{\frac{3k}{5(k-4)}}$, \, $E_{\rm K,iso,52} = E_{\rm K,iso}/10^{52}$, \, $\epsilon_{e,-1}=\epsilon_e/0.1$, \, $\epsilon_{B,-2}=\epsilon_B/0.01$,\, and\, $t_5=t/10^5$.
}. For $k=0$ and 2, our results return to those given by \citet[see Table 13 therein]{Gao+2013}. 

%In the regime  $\nu<\nu_{\rm a}<\nu_{\rm m}<\nu_{\rm c}$, one has
%\begin{equation}
%F_{\nu} =F_{\nu, {\rm max}}  \left(\frac{\nu_{\rm a}}{\nu_{\rm m}} \right)^{\frac{1}{3}}  \left(\frac{\nu}{\nu_{\rm a}} \right)^2  \propto t^{\frac{2}{4-k}} \nu^2    
%\end{equation}

%In the regime  $\nu_{\rm a}<\nu<\nu_{\rm m}<\nu_{\rm c}$, one has
%\begin{equation}
%F_{\nu} =F_{\nu, {\rm max}} \left( \frac{\nu}{\nu_{\rm m}}  \right)^{\frac{1}{3}} \propto t^{\frac{2-k}{4-k}} \nu^{\frac{1}{3}}    
%\end{equation}

%In the regime  $\nu_{\rm a}<\nu_{\rm m}<\nu<\nu_{\rm c}$, one has
%\begin{equation}
%F_{\nu} =F_{\nu, {\rm max}} \left( \frac{\nu}{\nu_{\rm m}}  \right)^{-\frac{p-1}{2}} \propto t^{-\frac{12p-12+5k-3kp}{16-4k}} \nu^{-\frac{p-1}{2}}    
%\end{equation}

%In the regime  $\nu_{\rm a}<\nu_{\rm m}<\nu_{\rm c}<\nu$, one has 
%\begin{equation}
%F_{\nu} =F_{\nu, {\rm max}} \left(\frac{\nu_{\rm c}}{\nu_{\rm m}} \right)^{-\frac{p-1}{2}}  \left(\frac{\nu}{\nu_{\rm c}} \right)^{-\frac{p}{2}} \propto t^{-\frac{3p-2}{4}} \nu^{-\frac{p}{2}}  .  
%\end{equation}

%%%%%%%%%%%%%%%%%%%%%%%%%%%%%%%%%%%%%%%%%%%%%%%%%%
\begin{table}
\caption{ Closure relations (CRs) for different CNM profile in the $\nu_a<\nu_m<\nu_c$ spectral regime.} \label{tab:CR}
\centering
\fontsize{8}{11}\selectfont    %{字体尺寸}{行距}
\begin{threeparttable}
\begin{tabular}{ccccc}
\toprule
\multirow{2}{*}{}& \multirow{2}{*}{}&\multirow{2}{*}{$\beta$}&\multirow{2}{*}{$\alpha$}&\multirow{2}{*}{$\alpha(\beta)$} \\
 & & & \\
\cmidrule(lr){1-5}%1-5列画横线
\multirow{4}{*}{k=0, slow cooling}&$\nu<\nu_a$&  -2&   $-\frac{1}{2}$&    $\frac{1}{4}\beta$ \\
& $\nu_a<\nu<\nu_m$&  $-\frac{1}{3}$&    $-\frac{1}{2}$&   $\frac{3}{2}\beta$\\
& $\nu_m<\nu<\nu_c$&  $\frac{p-1}{2}$&    $\frac{3(p-1)}{4}$&     $\frac{3}{2}\beta$\\
& $\nu_c<\nu$& $\frac{p}{2}$&   $\frac{3p-2}{4}$&    $\frac{3\beta-1}{2}$\\
\cmidrule(lr){1-5}
\multirow{4}{*}{k=1, slow cooling}&$\nu<\nu_a$& -2& $-\frac{2}{3}$&   $\frac{1}{3}\beta$\\
& $\nu_a<\nu<\nu_m$& $-\frac{1}{3}$&   $-\frac{1}{3}$&  $\beta$\\
& $\nu_m<\nu<\nu_c$& $\frac{p-1}{2}$&      $\frac{9p-7}{12}$&      $\frac{9\beta+1}{6}$\\
& $\nu_c<\nu$& $\frac{p}{2}$&  $\frac{3p-2}{4}$&    $\frac{3\beta-1}{2}$\\
\cmidrule(lr){1-5}
\multirow{4}{*}{k=1.5, slow cooling}&$\nu<\nu_a$& -2&  $-\frac{4}{5}$&   $\frac{2}{5}\beta$\\
& $\nu_a<\nu<\nu_m$& $-\frac{1}{3}$&   $-\frac{1}{5}$&    $\frac{3}{5}\beta$\\
& $\nu_m<\nu<\nu_c$& $\frac{p-1}{2}$&      $\frac{15p-9}{20}$&      $\frac{15\beta+3}{10}$\\
& $\nu_c<\nu$& $\frac{p}{2}$&  $\frac{3p-2}{4}$&    $\frac{3\beta-1}{2}$\\
\cmidrule(lr){1-5}
\multirow{4}{*}{k=2.0, slow cooling}&$\nu<\nu_a$& -2& -1&   $\frac{1}{2}\beta$\\
& $\nu_a<\nu<\nu_m$& $-\frac{1}{3}$&   0&  0\\
& $\nu_m<\nu<\nu_c$& $\frac{p-1}{2}$&      $\frac{3p-1}{4}$&      $\frac{3\beta+1}{2}$\\
& $\nu_c<\nu$& $\frac{p}{2}$&  $\frac{3p-2}{4}$&    $\frac{3\beta-1}{2}$\\
\cmidrule(lr){1-5}
\multirow{4}{*}{k=2.5, slow cooling}&$\nu<\nu_a$& -2&  $-\frac{4}{3}$&   $\frac{2}{3}\beta$\\
& $\nu_a<\nu<\nu_m$& $-\frac{1}{3}$&   $\frac{1}{3}$&  $-\beta$\\
& $\nu_m<\nu<\nu_c$& $\frac{p-1}{2}$&      $\frac{9p+1}{12}$&      $\frac{9\beta+5}{6}$\\
& $\nu_c<\nu$& $\frac{p}{2}$&  $\frac{3p-2}{4}$&    $\frac{3\beta-1}{2}$\\
\bottomrule
\end{tabular}\vspace{0cm}
\end{threeparttable}
\end{table}

The CRs for different CNM profile in the $\nu_a<\nu_m<\nu_c$ spectral regime are listed in Table \ref{tab:CR}. Inspecting the radio spectrum as given by \citet{Andreoni2022}, the low-frequency bands $< 20$ GHz should be in the $\nu<\nu_{\rm a}$ regime with spectral index $\beta=-2$ and temporal index $\alpha=-2/(4-k)$. The temporal index for 15.5 GHz is $\alpha\simeq -0.79$ obtained by \citet{Pasham2023}, which is close to the $k=1.5$ case ($\alpha=-4/5$) by inspecting Table \ref{tab:CR}. Applying the closure relation, i.e., equating $2/(4-k)=0.79$, one can find $k\simeq 1.47$, preferring a bondi-like profile. This result is also consistent with the equipartition analysis result of \citet{Matsumoto2023}, and is similar to that implied by modeling the early radio emission from the first jetted TDE Sw J1644+57 \citep{Metzger2012,Berger2012}.

\section{Constraining the CNM Profile with FS Model Fit}\label{sec:model}
%\subsection{Afterglow Model Fit}
To get a better constraint on the CNM profile index $k$, we employ a numerical code \texttt{PyFRS}\footnote{\url{https://github.com/leiwh/PyFRS}} for FS model described in \citet{Wang2014}, \citet{Lei2016} and \citet{Zhu2023}. The dynamical evolution of the shell is calculated numerically using a set of hydrodynamical equations \citep{Huang2000}
\begin{equation}
\frac{dR}{dt} =\beta_{\rm j} c \Gamma (\Gamma+\sqrt{\Gamma^2 -1}),    
\end{equation}
\begin{equation}
\frac{dm}{dR} =2\pi R^2 (1-\cos\theta_{\rm j})n m_{\rm p},   
\end{equation}
\begin{equation}
\frac{d\Gamma}{dm} = -\frac{\Gamma^2 -1}{M_{\rm ej} + 2\Gamma m} , 
\end{equation}
where $R$ and $t$ are the radius and time of the event in the source frame, $m$ is the swept-up mass, $M_{\rm ej}=E_{\rm K, iso} (1-\cos\theta_{\rm j})/2(\Gamma_0-1)c^2$ is the ejecta mass, and  $\beta_{\rm j}=\sqrt{\Gamma^2-1}/\Gamma$ . 
The density profile is described by a power-law in radius $R$ as Equation (\ref{eq:CNM_profile}). The synchrotron spectra of the jet is calculated following the standard broken-power-law spectral model separated by the three characteristic frequencies ($\nu_{\rm a},\nu_{\rm m}, \nu_{\rm c}$) as described in Section \ref{sec:CR} \citep[for a detailed review]{Gao+2013, Zhang2018}.

Now, we can numerically calculate the synchrotron light curve and spectrum from the forward shock for different CNM density profile with our \texttt{PyFRS} code. Markov chain Monte Carlo (MCMC) with \texttt{PyFRS} is adopted for multi-band fitting to place constraints on the model parameters. The MCMC fitting is done using the python package \textit{emcee} \citep{emcee2013}, which utilizes a group of parallel-tempered affine invariant walkers to explore the parameter space.

Eight free parameters are considered in the fitting, i.e., the isotropic kinetic energy $E_{\rm K,iso}$, the initial lorentz factor $\Gamma_0$ of the jet, the jet opening angle $\theta_{\rm j}$, the CNM profile parameter $k$, the number density of CNM medium $n_{18}$ at $R=10^{18} \rm cm$, the electron distribution power-law index $p$, the energy fraction in electrons $\epsilon_{\rm e}$ and in magnetic field $\epsilon_{\rm B}$. The viewing angle $\theta_{\rm obs}$ is fixed to $0^\circ$ in the fit due to the on-axis jet as suggested by \citet{Andreoni2022} and \citet{Pasham2023}, and also by the nearly zero observed polarization degrees \citep{Cikota2023}. 

The prior range of these parameters are given based on the estimations from observations. First, the CR studies in Section \ref{sec:CR} suggest the CNM density profile index of $k\simeq 1.5$, favoring a Bondi-like profile. However, other CNM density profile types ($k=1$ or $5/2$) can not be completely ruled out by such a CR study. We therefore set the prior range of $k$ as $1 - 3$. 

AT2022cmc is bright in X-ray with peak luminosity $\geq 3\times 10^{47} \rm erg \ s^{-1}$, which is comparable with the well-studied relativistic jetted TDE Sw J1644+57 \citep{Andreoni2022,Matsumoto2023}. The X-ray light curve shows rapid decay as $L_{\rm X,iso} \simeq 3\times 10^{47} \rm erg \ s^{-1} (t/5 {\rm days})^{-2}$, indicating an isotropic equivalent energy of $E_{\rm X,iso}>10^{53} {\rm erg}$. The kinetic energy $E_{\rm K, iso}$ can be larger or smaller than this X-ray energy $E_{\rm X,iso}$ (depending on the efficiency of converting jet power to X-ray radiation), we take the prior range of isotropic kinetic jet energy $E_{\rm K, iso}$  as $10^{52} - 10^{54} {\rm erg}$. 

From the day-timescale variability of the radio data, \citet{Rhodes2023} infer that the bulk Lorentz factor of the jet is $\, \gtrsim 8$. By extrapolation of the equipartition analysis results, \citet{Matsumoto2023} obtained $\Gamma_0 \simeq 5$ for a narrow jet model or $2.5$ for a wide jet model. Using the synchrotron afterglow model of the relativistic jet, a Lorentz factor of $12$ was obtained by \citet{Andreoni2022}. In \citet{Pasham2023}, the bulk Lorentz factor of the jet is $86$ for the emission model with only synchrotron and synchrotron self-Compton (SSC), and $5$ for the model with external Compton (EC) included. The prior range $1.1 - 50 $ is used for $\Gamma_0$ in the MCMC fit. 

The jet opening angle is poorly constrained from observations. It is reasonable to assume $\theta_{\rm j} \sim 1/\Gamma_0 \sim 0.2$ (0.4) if $\Gamma_0 \simeq 5$ (2.5) is adopted. Another choice $\theta_{\rm j} \sim 0.1$ is motivated by the modeling of Sw J1644+57 \citep{Metzger2012}. We use prior range $0.052-0.35$ for $\theta_{\rm j}$.

Taking the peak ($\sim 1$ day) of the early optical phase (coincides with the peak of X-ray) as the onset of deceleration phase, one finds the  deceleration time in the engine's rest frame as $t_{\rm dec}\simeq 1 \ {\rm day}/(1+z) \sim 0.5\ {\rm day}$  \citep{Matsumoto2023}. The deceleration radius is given by $R_{\rm dec}\simeq 2\Gamma_0^2 c t_{\rm dec}\sim 6.5\times 10^{16} {\rm cm} \ (\Gamma_0/5)^2 (t_{\rm dec}/0.5\ {\rm day})$. We thus can estimate the jet kinetic energy $E_{\rm K, iso} \simeq \frac{4\pi}{3-k} m_{\rm p} c^2 \Gamma_0^2 R_{\rm dec}^3 n(R_{\rm dec})$, where $n(R_{\rm dec})$ is the density at $R_{\rm dec}$. The CNM density at $R=10^{18} \rm cm$ can thus be given by

\begin{equation}
    n_{18}\simeq 19\,{\rm cm^{-3}} \left(\frac{E_{\rm K, iso}}{10^{53} {\rm erg}} \right)  \left(\frac{\Gamma_0}{5} \right)^{-5}  \left(\frac{t_{\rm dec}}{0.5\ {\rm day}} \right)^{-1.5},
\end{equation}
where $k=1.5$ is taken as suggested by our CR studies in Section \ref{sec:CR}. We use a relatively wide prior range $10^{-3} - 200\ \rm cm^{-3}$ for $n_{18}$ by considering the uncertainty in the above estimations.

The multi-band fit to early optical non-thermal component finds a spectral index $\beta=-1.32\pm 0.18$ \citep{Andreoni2022}, leading to an electron energy index of $p=2.64\pm 0.36$ \citep{Matsumoto2023}. We thus fix $p$ to 2.64 in the fit.

%For the two energy fractions $\epsilon_{\rm e}$ and $\epsilon_{\rm B}$, we adopt the typical value ranges. However, 
The low polarization (consistent with zero) favors the jet with low magnetic field energy density \citep{Cikota2023}. We therefore use a relatively smaller value of the lower-limit for $\epsilon_{\rm B}$. The descriptions, prior type, and prior range for each model parameter are presented in Table \ref{tab:fit}.

%By applying different models on AT2022cmc, different Lorentz factor\,(LF) of the jet are obtained. Using the synchrotron afterglow model of the relativistic jet, a LF of $12$ was obtained by \citet{Andreoni2022}. In the synchrotron and SSC models, the bulk Lorentz factor of the jet is $86$. In the external compton (EC) model, the bulk Lorentz factor of the jet is $5$ \citep{Pasham2023}. From the day-timescale variability of the radio data, \citet{Rhodes2023} infer that the bulk Lorentz factor of a jet that is pointing directly towards Earth is$\, \gtrsim 8$.
% Considering the CNM density distribution, \citet{Matsumoto2023} employed the equipartition analysis and afterglow model on AT2022cmc roughly estimated the parameters of the jet as below (see table 2 of \citet{Matsumoto2023}):  the initial LF is $\Gamma \sim 5 $, the isotropic energy is $E_{\rm j,iso} \sim 4 \times 10^{53}\,\rm{erg} $, the opening angle is $\theta_j \sim 0.15$, the distribution slope of the electrons' energy is $p\sim 2.9$, the energy fraction of non-thermal electrons and magnetic filed is 0.2 and 0.002, respectively, the density of external medium at $10^{17}\rm cm$ is $200\,\rm cm^{-3}$, and the CNM profile parameter is $k\sim 1.8$ preference the bondi-accretion like Sw J1644+57

The X-ray light curves are highly variable (on timescales of 1000 s), which are assumed to arise from the internal emission of the jet as in Sw J1644+57 \citep{Matsumoto2023}. The optical and UV observations revealed a fast-fading red ``flare'' ($<5$ days) with a peak luminosity $\nu L_\nu \simeq 10^{46} \rm{erg\ s^{-1}}$ that transitioned quickly to a slow blue ``plateau'' of luminosity $\sim 10^{45} \rm{erg\ s^{-1}}$ lasting at least a couple of months, suggesting a two-component emission model \citep{Andreoni2022}: jet component (early r-band) and accretion disk thermal component (blue optical plateau with a blackbody temperature of $\simeq (2-4)\times 10^4$ K). The late-time r-band ($\geq 5$ days) data likely have a thermal origin similar to other optically-selected TDEs. The radio counterpart was identified in Karl G. Jansky Very Large Array (VLA) observations on 15 February 2022 (four days after the ZTF trigger).
%Four days after ZTF trigger time, On 2022 February 15, the radio emission was identified in VLA.
The radio/sub-millimeter observations display a typical synchrotron self-absorption spectrum \citep{Andreoni2022,Pasham2023}. As studied in \citet{Andreoni2022} and \citet{Matsumoto2023}, the radio, sub-millimeter, and early r-band optical emissions should originate from the jet-CNM interactions. Therefore, for our MCMC fitting with \texttt{PyFRS}, we use the sub-millimeter/radio (7.0 GHz, 8.5 GHz, 10.5 GHz, 11.5GHz, 15.5 GHz, 17.4GHz, 33.5 GHz, 86 GHz, 102 GHz, 225 GHz, and 350 GHz), and the early r-band optical ($< 5$ days) data. 

\begin{figure*}[htp]
\center
\includegraphics[width=0.9\textwidth]{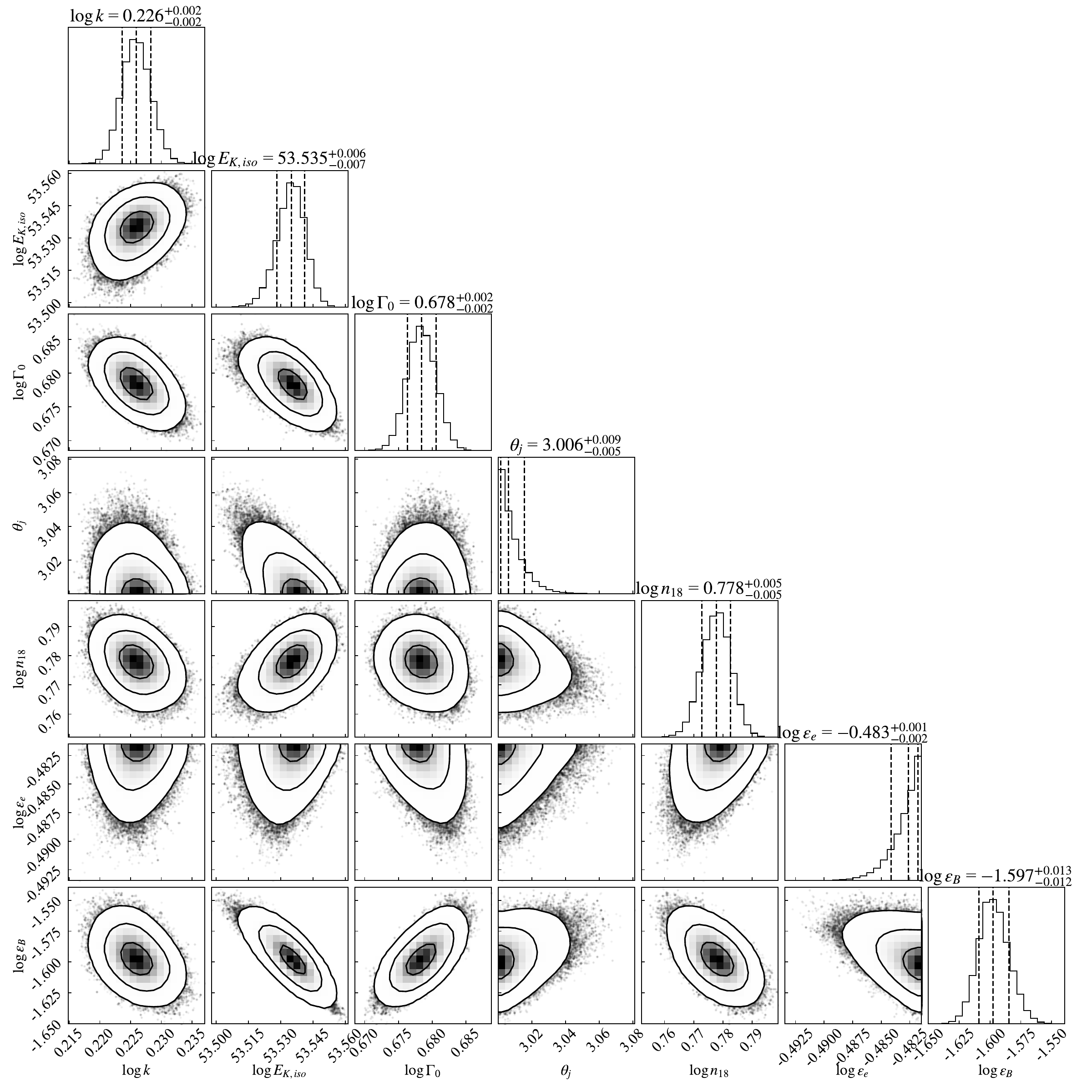}
\caption{Posterior distribution and parameter constraints were obtained using FS modelling of AT2022cmc with MCMC. The median values with the $1\sigma$ error regions are also shown in the one-dimensional probability distribution.}
\label{fig:model_corner}
\end{figure*}

\begin{figure}[htp]
\center
\includegraphics[width=0.8\textwidth]{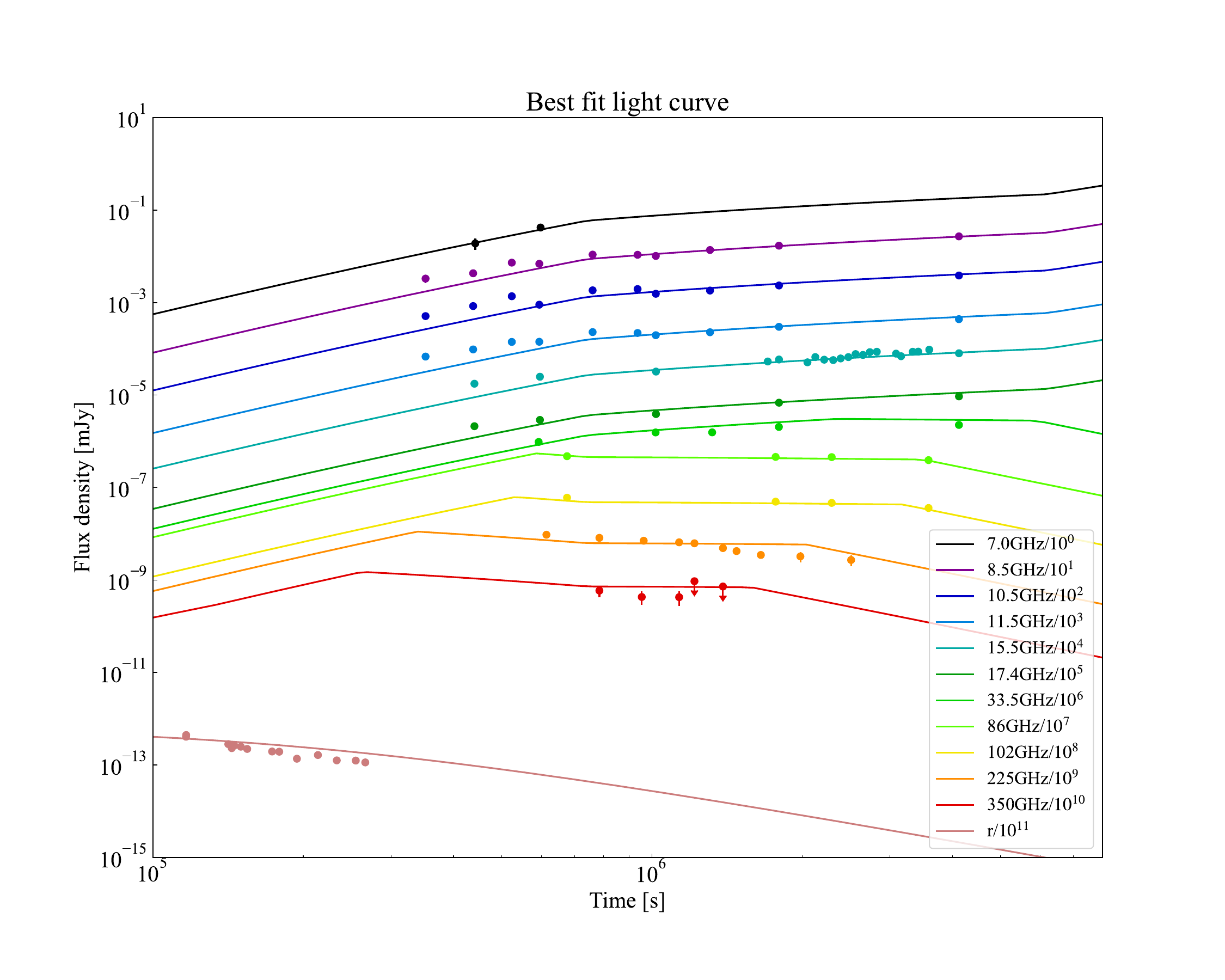}
\caption{Optical, sub-millimeter, and radio afterglow data of AT2022cmc along with the best fit with FS code. Bands are in different colors with shift factors in legend.} 
\label{fig:model_lc}
\end{figure}

We performed a parameter search with 64 walkers over 20000 iterations, discarding the first 10000 as burn-in steps. The posterior distribution of the model parameters are shown in Fig.~\ref{fig:model_corner}. The best fit of each parameter is given in Table~\ref{tab:fit} as: $k=1.68$, $n_{18} = 6.00 \ {\rm cm^{-3}}$, $E_{\rm K,iso}=3.43\times 10^{53}$ erg, $\Gamma_0=4.76$, $\theta_{\rm j}=5.25\times10^{-2}$ rad, $\epsilon_{\rm e}=3.29\times 10^{-1}$ and $\epsilon_{\rm B}=2.53\times 10^{-2}$. In Fig.~\ref{fig:model_lc}, we have shown the optical, sub-millimeter, and radio afterglow light curves of AT2022cmc along with the best-fit model. The CNM type constrained with our fit is consistent with the CR analysis in Section\,\ref{sec:CR}, both indicating a Bondi-like profile.

\begin{figure*}[htp]
\center
\includegraphics[width=0.8\textwidth]{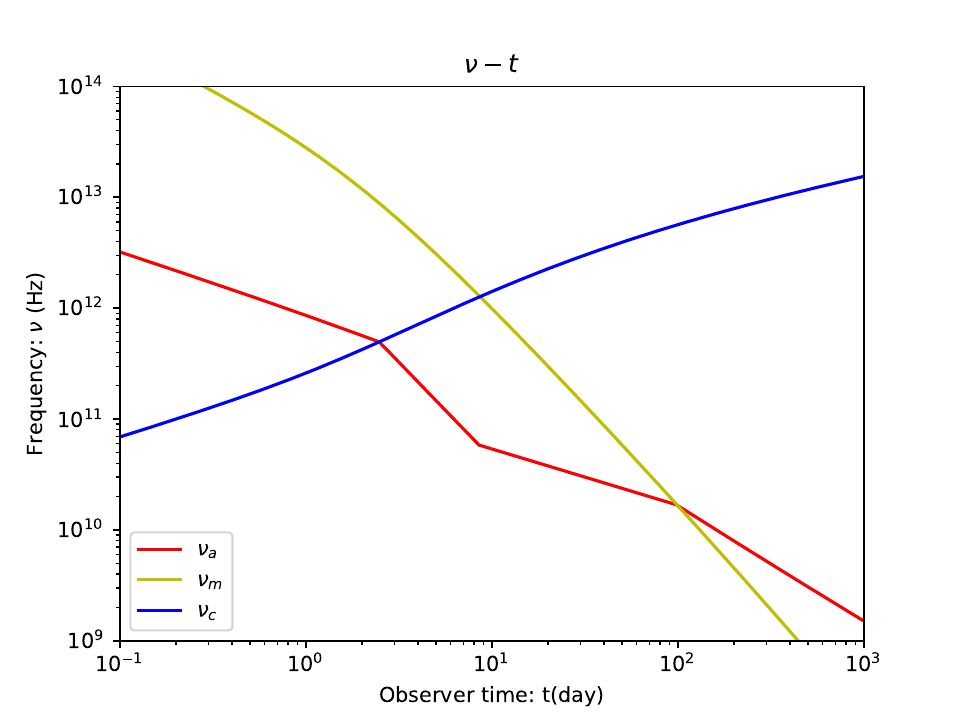}
\caption{The time-evolution of the characteristic synchrotron frequencies ( $\nu_a$, $\nu_m$ and $\nu_c$) based on our fitting results.}
\label{fig:nu}
\end{figure*}

In Fig.~\ref{fig:nu}, we present the time-evolution of the characteristic synchrotron frequencies ( $\nu_a$, $\nu_m$ and $\nu_c$) based on our best-fit results (as shown in Table~\ref{tab:fit}). One can see that the main observations of 15.5 GHz are located in the $\nu_a < \nu_m < \nu_c$ regime, which is adopted by the CR analysis in Section\,\ref{sec:CR}. 

\begin{table}[htbp]
\begin{center}
\caption{The input parameters, prior type, prior range, best-fit value of multi-band modelling of AT2022cmc.}
\label{tab:fit}
\begin{tabular}{lrrrr}
\hline\hline
     Parameters &  Prior Type & Prior Range  & Best fit  &Uncertainty \\  
    \hline
    $k$ & log flat & 1-3 & 1.68  &  $(+0.008,-0.008)$ \\
    $n_{18}\rm\ (cm^{-3})$   & log flat & $10^{-3}-200$ & 6.00  & $(+0.069,-0.069)$  \\ 
    $E_{\rm K, iso}$ (erg) & log flat & $10^{52}- 10^{54}$ & $3.43\times10^{53}$ & $(+4.736\times10^{51},-5.525\times10^{51})$ \\
    $\Gamma_0$ & log flat & 1.1-50 & 4.76  & $(+0.022,-0.022)$ \\    
    $\theta_{\rm j}$ (rad) & flat & $0.052-0.35$ & $5.25\times10^{-2}$    &$(+1.571\times10^{-4},-0.873\times10^{-4})$   \\ 
    %$p$   & flat & $2.1-3.0$ & 2.64 \\ 
    $\epsilon_{\rm e}$   & log flat & $10^{-3}-0.33$ &  $3.29\times10^{-1}$   & $(+0.757\times10^{-3},-1.514\times10^{-3})$   \\ 
    $\epsilon_{\rm B}$   & log flat & $10^{-4}-0.33$ &  $2.53\times10^{-2}$    & $(+0.757\times10^{-3},-0.699\times10^{-3})$  \\ 
    \hline
\end{tabular}
\end{center}
\end{table}

%\begin{figure*}[htp]
%\R{For convenience, we added offsets in fluxes as shown in the figure legend.}

%\tablecomments{This is an example of how to split a deluxetable. You can split any table with this command into two or three parts.  The location of the split is given by the author based on the placement of the ``B'' indicators in the column identifier preamble.  For more information please look at the new \aastex instructions.}

\section{Discussion} \label{sec:disussion}
\subsection{Accretion History of the SMBH in AT2022cmc}
The pre-existing CNM density profile $n=n_{18} (R/10^{18} {\rm cm})^{-k}$ probed by the FS can provide clues to star formation activity locally in the star cluster or accretion history of central SMBH. 

First, an upper limit on SMBH mass of AT2022cmc $M_\bullet<4.7\times 10^8 M_\odot$ is obtained by using the galaxy bulge - black hole mass relation and the upper limit on the AT2022cmc galaxy mass ($<10^{11.2} M_\odot$) \citep{Andreoni2022}. The X-ray light curve shows a variability timescale of $\delta t\simeq 1000/(1+z) {\rm s} \simeq 456 \rm s$. Assuming the variability timescale is defined by the marginally stable orbit of the accretion disk $R_{\rm ms}/c$, the SMBH mass can be estimated as 
\begin{equation}
    M_\bullet\simeq 1.8\times 10^7 M_\odot \left(\frac{R_{\rm ms}}{5 R_{\rm g}} \right)^{-1} \left(\frac{\delta t}{456 \rm s}\right),
\end{equation}
where $R_{\rm g}=GM_\bullet/c^2$, and $R_{\rm ms} \simeq 6 R_{\rm g}$ ($1 R_{\rm g}$) for SMBH spin $a_\bullet =0$ (1). Under assumption that the jet is powered by Blandford-Znajek mechanism\citep{BZ77,Lei2011,Liu2015,Liu2017}, one obtains $a_\bullet \geq 0.3$ for the beaming-corrected peak jet power $L_{\rm jet}=3\times 10^{46} {\rm erg\ s^{-1}}$\citep{Andreoni2022}. Thus, $a_\bullet = 0.3$ is adopted in the estimations ($R_{\rm ms}(a_\bullet=0.3)\simeq 5 R_{\rm g}$), and we have $M_{\bullet}< 1.8\times 10^7 M_\odot$. On the other hand, given that the TDE thermal optical emission originates from a quasi-spherical hydrostatic envelope radiating near the SMBH Eddington limit, the observed optical plateau luminosity $\sim 10^{45} \rm{erg\ s^{-1}}$ of AT2022cmc would require a SMBH mass $M_{\bullet} \sim 10^7 M_\odot$.

Our best fit result $k=1.68$ (see Section\,\ref{sec:model}) is close to the Bondi accretion prediction ($k=1.5$) \citep{Quataert1999}. Using the best fit result $n_{18}\simeq 6.00\ \rm cm^{-3}$, we obtain the mass accretion rate of $\dot{M}_{\rm acc} \simeq 5.1\times 10^{-4} \dot{M}_{\rm Edd} m_{\bullet,7}^{-1/2}$, where $\dot{M}_{\rm Edd}$ is the Eddington accretion rate, and $m_{\bullet,7}\equiv M_\bullet/10^7M_\odot$ is the mass of SMBH in units of $10^7M_\odot$. This would support the nucleus of AT2022cmc being a low-luminosity active galactic nucleus (AGN). The total disk luminosity of this AGN will be (assuming a thin disk model)
\begin{equation}
    L_{\rm disk} = (1-E_{\rm ms}) \dot{M}_{\rm acc} c^2 \sim 4.4\times 10^{41} {\rm erg\ s^{-1} } m_{\bullet,7}^{1/2},
\end{equation}
where $E_{\rm ms}$ is the specific energy corresponding to the radius of the marginally stable orbit $R_{\rm ms}$ \citep{Bardeen1972}, and $1-E_{\rm ms} \simeq 0.06$ (0.42) for SMBH spin $a_\bullet =0$ (1). In Equation (16), $a_\bullet = 0.3$ is adopted in the estimation of AGN disk luminosity. This AGN luminosity $L_{\rm disk} \sim 4.4\times 10^{41} {\rm erg\ s^{-1} }$ (if $M_{\bullet} \sim 10^7 M_\odot$) is much weaker than the observed long-lasting optical ``plateau'' luminosity ($\sim 10^{45} \rm{erg\ s^{-1}}$) of AT2022cmc.

It should be noted that the medium density around SMBH might be asymmetric. The radio emission from outflow may provide additional information to study the complex structure, e.g., the cloud \citep{Mou2022, Perlman2022, Bu2023} and torus \citep{Mou2021}. The surrounding dust distribution at comparable distances can also be indirectly probed via the detection of infrared dust echoes \citep{Jiang2016}.

%The \textbf{instabilities} developed in the jet-CNM interactions may influence the afterglow emissions, which should be explored in future MHD simulations.

The instability developed in the jet-CNM interactions may influence the afterglow emission, which should be explored in future MHD simulations. Generally, as the jets produced by the Blandford-Znajek mechanism around spinning SMBHs propagate to $\sim$ sub-parsec or parsec scales, they are expected to undergo significantly instabilities, primarily by the current-driven kink instability (e.g., \citet{Guan2014}; \citet{Ressler2021}; \citet{Lalakos2023}). Given that in our model fitting, the FS and jet will propagate to $\sim 0.8$ parsec, the interaction between the jet and the CNM is expected to be in 3D and likely unstable. 
%, with densities lower in the jet funnel and higher in the direction close to the disk. 

%infrared observations of the dust emission from the host nucleus, which reveal a light echo from the flare ()

%\subsection{Effects of Jet-CNM Instability}

\subsection{Two-component Jet Model Fit}
Inspecting Fig.~\ref{fig:model_lc}, we find that the late-time data of AT2022cmc is generally consistent with the FS model predictions. The early light curves ($t<8$ days) are, however, poorly reproduced with this one-component jet model. Modeling the multi-wavelength radio light curves of the well-studied jetted TDE Sw J1644+57 implied a two-component jet structure \citep{Wang2014,LiuDangbo2015,Mimica2015}. The radio spectrum study of AT2022cmc by \citet{Matsumoto2023} suggested an extra jet component or energy injection. Motivated by these studies, we then try to fit the multi-band light curves of AT2022cmc with a two-component jet model developed in \citet{Wang2014} but modified to adapt to an arbitrary CNM density profile. The calculation for the FS is still based on the \texttt{PyFRS} code.
%We also modified the code to adapt to an arbitrary CNM density profile.    

\begin{figure}[htp]
\center
\includegraphics[width=0.9\textwidth]{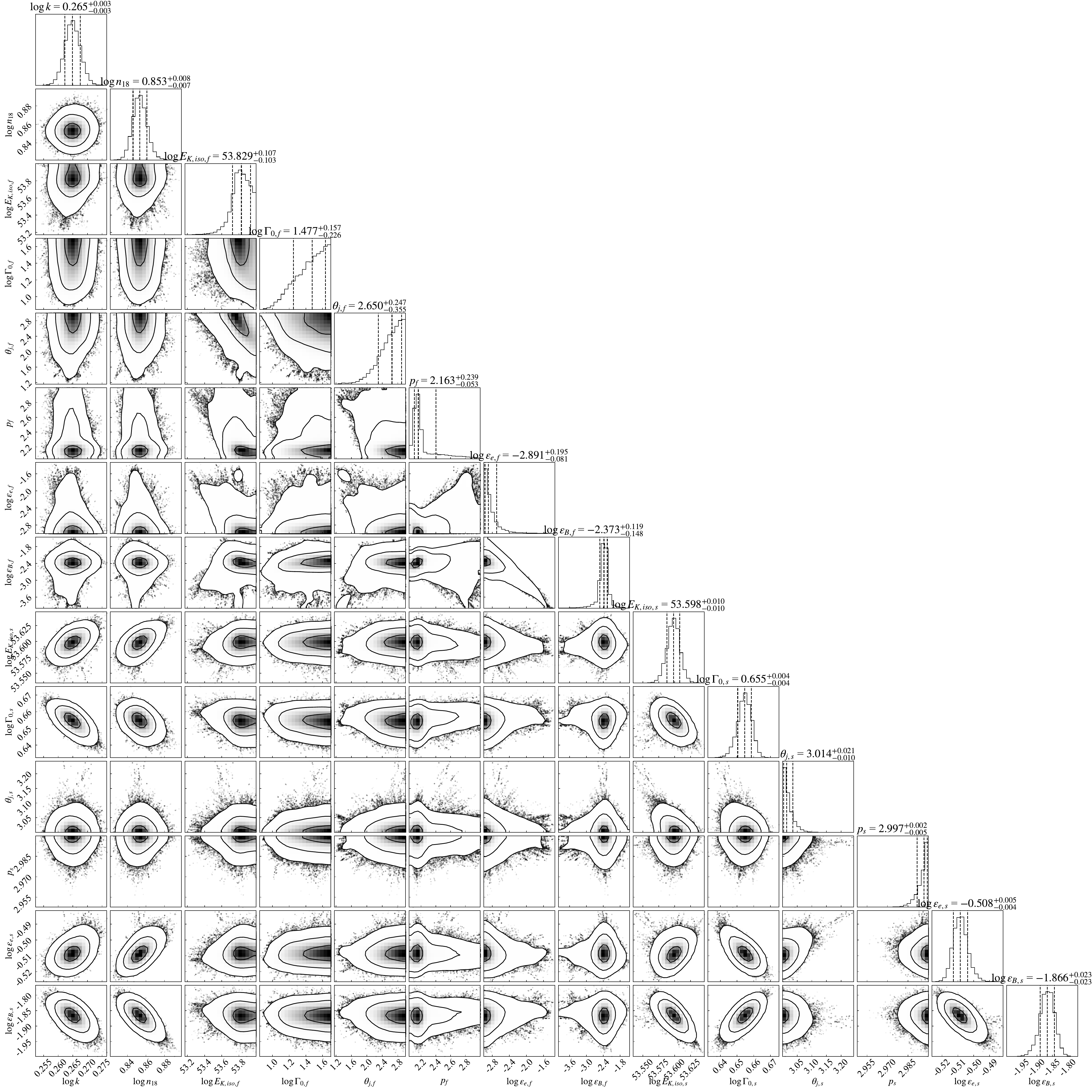}
\caption{Posterior distribution and parameter constraints were obtained using a two-component jet model. The median values with the $1\sigma$ error regions are also shown in the one-dimensional probability distribution.}
\label{fig:model_corner_2comp}
\end{figure}

As in Section\,\ref{sec:model}, the MCMC fit with the two-component jet model is also performed with 64 walkers over 20000 iterations. The posterior distribution of the model parameters are shown in Fig.~\ref{fig:model_corner_2comp}. We find a fast-component jet with initial Lorentz factor of $\Gamma_{0,\rm f} =29.99$, $E_{\rm K,iso,f}=6.75\times 10^{53}$ erg, $\theta_{\rm j,f}=4.63\times10^{-2}$ rad, $p_{\rm f} = 2.16$, $\epsilon_{\rm e,f}=1.29\times 10^{-3}$ and $\epsilon_{\rm B,f}=4.24\times 10^{-3}$, and a slow-component with initial Lorentz factor of $\Gamma_{0,\rm s} =4.52$, $E_{\rm K,iso,s}=3.96\times 10^{53}$ erg, $\theta_{\rm j,s}=5.26\times10^{-2}$ rad, $p_{\rm s} = 3.00$, $\epsilon_{\rm e,s}=3.10\times 10^{-1}$ and $\epsilon_{\rm B,s}=1.36\times 10^{-2}$. We obtain a CNM environment of AT2022cmc characterized by parameters $k=1.84$ and $n_{18} = 7.13 \ {\rm cm^{-3}}$. The best-fit values of the parameters are given in Table\,\ref{tab:fit-2comp}. The subscripts ``f '' and ``s'' denote the fast and slow components, respectively. 

%\ref{tab:fit} as: $E_{\rm K,iso}=3.47\times 10^{53}$ erg, $\theta_{\rm j}=0.10$ rad, $n_{18} = 6.77 \ {\rm cm^{-3}}$, $p = 2.64$, $\epsilon_{\rm e}=0.33$ and $\epsilon_{\rm B}=2.53\times 10^{-2}$. 

\begin{figure}[htp]
\center
\includegraphics[width=0.8\textwidth]{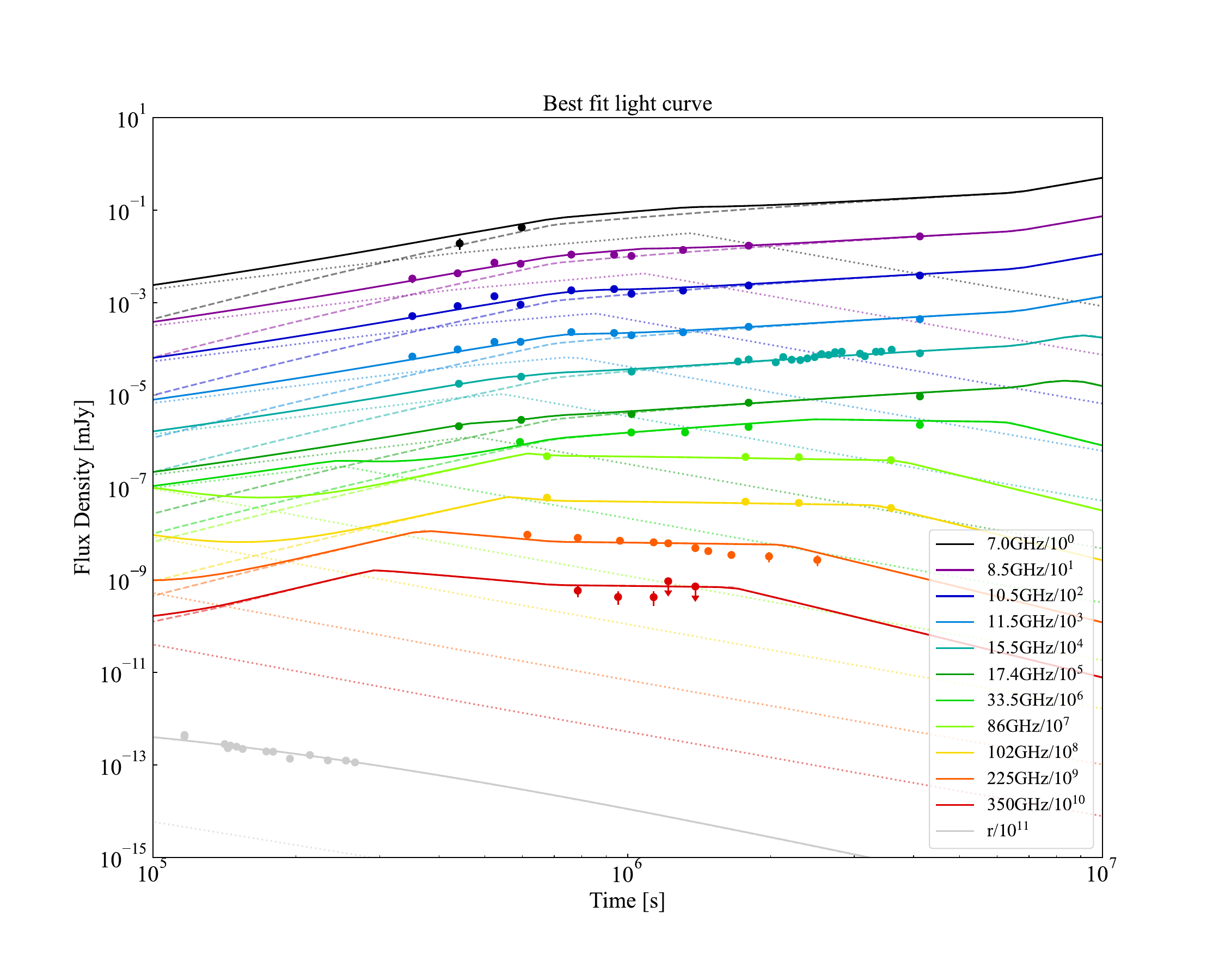}
\caption{Optical, sub-millimeter, and radio afterglow data of AT2022cmc along with the best fit with a two-component jet (a fast component with dotted lines, and a slow component with dashed lines) model. Bands are in different colors with shift factors in legend.} 
\label{fig:model_lc_2comp}
\end{figure}

The best-fit results for the optical, sub-millimeter, and radio afterglow light curves are presented in Fig.~\ref{fig:model_lc_2comp}. The fast and slow components are plotted with dotted and dashed lines, respectively. As one can see, the two-component jet model can well interpret both the early (by the combination of the fast and slow components) and late time data (dominated by the slow component see the dashed lines in Fig.~\ref{fig:model_lc_2comp}) of AT2022cmc. The result in Section\,\ref{sec:model} may correspond to the slow component in Fig.~\ref{fig:model_lc_2comp}. The constraint CNM density profile index $k\sim 1.8$ is also consistent with the one-component jet model ($k\sim 1.7$) in Section\,\ref{sec:model}. Therefore, our results support that AT2022cmc may contain a two-component jet as in Sw J1644+57. These two jetted TDEs (AT2022cmc and Sw J1644+57) may share similar jet physics.

%\begin{figure*}[htp]

\begin{table}[htbp]
\begin{center}
\caption{The input parameters, prior type, prior range, best-fit value of two-component jet model of AT2022cmc.}
\label{tab:fit-2comp}
\begin{tabular}{lrrrr}
\hline\hline
     Parameters &  Prior Type & Prior Range  & Best fit &Uncertainty \\  
    \hline
    $k$ & log flat & 1-3 & 1.84   &$(+0.013,-0.013)$  \\
    $n_{18}\rm\ (cm^{-3})$   & log flat & $10^{-5}-10^{3}$ & 7.13  &$(+0.131,-0.115)$ \\ 
    $E_{\rm K, iso,f}$ (erg) & log flat & $10^{49}- 10^{54}$ & $6.75\times10^{53}$  &  $(+1.662\times10^{53},-1.560\times10^{53})$\\ 
    $\Gamma_{0,\rm f}$ & log flat & 1.0-50 & 29.99  & $(+10.842,-15.607)$\\   
    $\theta_{\rm j,f}$ (rad) & flat & $0.052-0.35$ & $4.63\times10^{-2}$   & $(+4.311\times10^{-3},-6.196\times10^{-3})$ \\ 
    $p_{\rm f}$   & flat & $2.0-3.0$ & 2.16   &$(+0.239,-0.053)$ \\ 
    $\epsilon_{\rm e,f}$   & log flat & $10^{-3}-0.33$ &  $1.29\times10^{-3}$   & $(+0.577\times10^{-3},-0.240\times10^{-3})$ \\ 
    $\epsilon_{\rm B,f}$   & log flat & $10^{-4}-0.33$ &  $4.24\times10^{-3}$   &$(+1.161\times10^{-3},-1.444\times 10^{-3})$  \\ 
    $E_{\rm K, iso,s}$ (erg) & log flat & $10^{49}- 10^{54}$ & $3.96\times10^{53}$   & $(+9.125\times10^{51},-9.125\times10^{51})$ \\ 
    $\Gamma_{0,\rm s}$ & log flat & 1.1-50 & 4.52  &$(+0.042,-0.042)$ \\   
    $\theta_{\rm j,s}$ (rad) & flat & $0.052-0.35$ & $5.26\times10^{-2}$    & $(+0.367\times10^{-3},-0.175\times10^{-3})$\\ 
    $p_{\rm s}$   & flat & $2.0-3.0$ & 3.00 & $(+0.002,-0.005)$ \\ 
    $\epsilon_{\rm e,s}$   & log flat & $10^{-3}-0.33$ &  $3.10\times10^{-1}$    & $(+3.574\times10^{-3},-2.859\times10^{-3})$\\ 
    $\epsilon_{\rm B,s}$   & log flat & $10^{-4}-0.33$ &  $1.36\times10^{-2}$    & $(+0.721\times10^{-3},-0.721\times10^{-3})$\\ 
    \hline
\end{tabular}
\end{center}
\end{table}

%\begin{figure*}[htp]
%\end{figure*}

%\R{For convenience, we added offsets in fluxes as shown in the figure legend.}

\subsection{Detectability of AT2022cmc-like Events with Einstein Probe}

AT2022cmc is unique and is the first on-axis jetted TDE discovered in optical. It is also the furthest jetted TDE discovered to date. Future optical surveys, e.g., Large Synoptic Survey Telescope (LSST) and Wide Field Survey Telescope (WFST), have great potential to discover more AT2022cmc-like events \citep{LSST2020,WFST2023}. WFST is a photometric surveying facility located in western China, it has a 2.5-meter diameter primary mirror. The wide field survey (WFS) and the deep high-cadence survey (DHS) programs have been scheduled, covering a sky area of 8000 and 1000 square degrees, respectively. The future WFST surveys will certainly promote our understanding of the optical emission of TDEs \citep{Lin2022}.  However, X-ray observations might still be the key to the confirmation of a jetted TDE from the optically selected candidates. 

The Einstein Probe (EP) is a mission dedicated to time-domain astronomy to monitor and characterize the soft X-ray transient sky \citep{EP2015}. It is proposed by the Chinese Academy of Sciences (CAS). It is equipped with the Wide X-ray Telescope (WXT, 0.5-4keV) and the Follow-up X-ray Telescope (FXT, 0.3-10keV). WXT employs the lobster-eye micro-pore optics, the sensitivity is $0.26\times 10^{-10}\rm erg s^{-1} cm^{-2}$ for 1000-second exposure \citep{Yuan22}. A Wolter-I nested telescope is adopted by FXT, and the sensitivity of FXT can achieve the order of $10^{-14}\rm erg s^{-1} cm^{-2}$ with a 25-minute exposure \citep{Zhang2022}. EP is an ideal instrument to systematically search and study TDEs. It will be able to detect more relativistic TDEs with relativistic jet out to redshift $z>1$. 

\begin{figure}
\center
\includegraphics[width=0.8\textwidth]{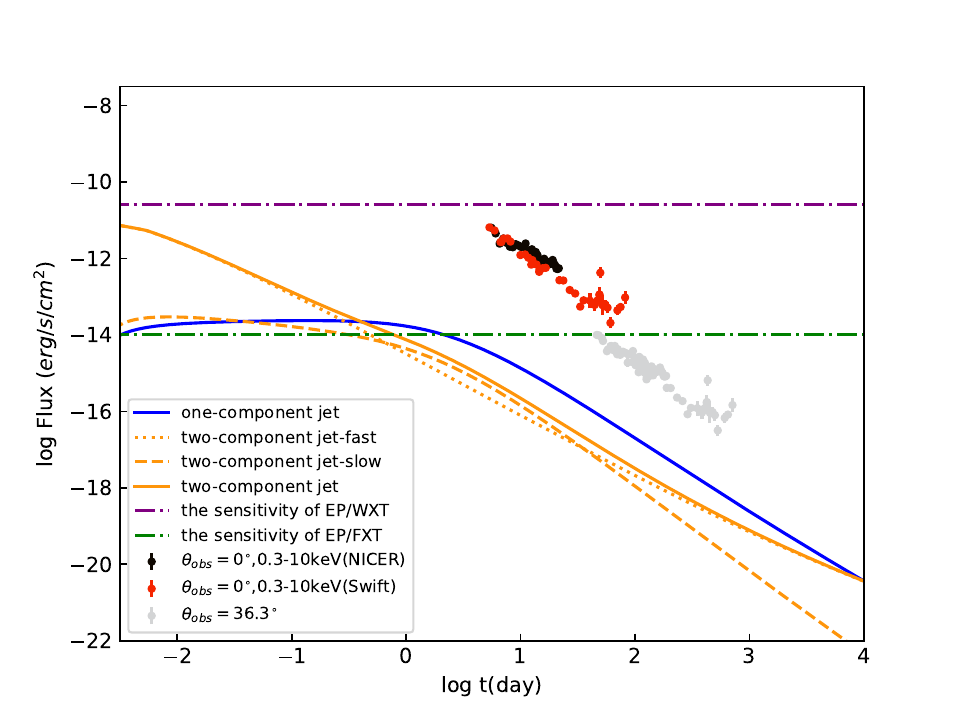}
\caption{X-ray light curves of AT2022cmc. The black (NICER) and red ({\em Swift}) points are the X-ray observations of AT2022cmc adopted from \citet{Pasham2023}. The blue solid line is the X-ray emission from the forward shock (FS) with the best-fit result of the one-component jet model (see Table~\ref{tab:fit}). The X-ray emission from the FS of the best-fit two-component jet model (see Table\,\ref{tab:fit-2comp}) is represented with yellow lines, in which the fast component, the slow component, and the combination of the two components are plotted with yellow dotted line, yellow dashed line, and yellow solid line, respectively. The sensitivity of the WXT and FXT of EP are also shown with purple and green dotted-dashed lines, respectively. The gray points are the observational X-ray data viewed off-axis with $\theta_{\rm obs} = 36.3^{\circ}$, which can be marginally detected by EP/FXT. }
\label{fig:x-ray_ep}
\end{figure}

In Fig.~\ref{fig:x-ray_ep}, we present the NICER (black points) and {\em Swift} (red points) X-ray data of AT2022cmc and the FS model predicted X-ray afterglow (solid lines), as well as the EP/WXT and EP/FXT sensitivity (dotted-dashed lines). It is found that AT2022cmc-like TDEs can be well monitored by EP/FXT, but are slightly below the EP/WXT sensitivity. Considering that the peak X-ray luminosity of AT2022cmc is larger than $3\times 10^{47} \rm erg \ s^{-1}$, such events still have good opportunities to be triggered in X-ray by EP/WXT.

As pointed out by \citet{Lei2016} and \citet{Yuan2016}, a good fraction of jetted TDEs might be viewed off-axis. The observed flux density is further subject to a correction factor due to the viewing angle for an off-axis observer \citep{Granot2002,Huang2021}
\begin{equation}
F_{\nu}(\psi,t) = a_{\rm off}^3 F_{\nu /a_{\rm off}}(0,a_{\rm off} t),
\label{eq:Fvoff}
\end{equation}
where $\psi=\rm{max}(0,\theta_{obs}-\theta_j)$ is the angle between the 
near-edge of the jet and the observer, and 
\begin{equation}
a_{\rm off}=\frac{{\mathcal D}_{\rm off}}{{\mathcal D}_{\rm on}}=
\frac{1-\beta}{1-\beta \cos \psi}, 
\label{eq:aoff}
\end{equation}
is the ratio of the on-beam Doppler factor to the off-beam Doppler factor,
with $\beta=\sqrt{1-1/\Gamma^2}$.

We take AT2022cmc as a prototype and investigate the detectability by EP when assuming different view angles $\theta_{\rm obs}$, see the gray points in Fig.~\ref{fig:x-ray_ep}. One can find that AT2022cmc-like events can be followed by EP/FXT when $\theta_{\rm obs} \leq 36.3^{\circ}$. Therefore, the combination of observations by WFST and EP has a large chance of revealing AT2022cmc-like events in the future.

%Fig.~\ref{fig:x-ray_ep} exhibits the detect probability of AT2022cmc-like events by EP.

%, and is expected to detect them at the peaks of their X-ray flares out to at least a few hundred Mpc.

\section{Summary}
\label{sec:summary}
AT2022cmc was discovered as a luminous, rapidly evolving transient by the Zwicky Transient Facility \citep{Andreoni2022,Pasham2023}. We employed the CR and MCMC analysis to constrain the CNM density profile of AT2022cmc under the framework of the FS model. Our conclusions are summarized as follows:

1. The r-band optical light curve as well as the radio/sub-millimeter observations of AT2022cmc can be explained by the synchrotron emission from the forward shock of the jet-CNM interactions. The best-fit parameter values of the one-component jet model are given in Table~\ref{tab:fit}. The initial Lorentz factor of the jet is 4.76. The isotropic jet kinetic energy $E_{\rm K, iso} = 3.43\times 10^{53} \rm erg$ indicates a prompt radiative efficiency of the jet $\epsilon_{\rm X}\equiv E_{\rm X, iso}/(E_{\rm X, iso}+E_{\rm K, iso}) \simeq 0.23$ if the observed X-ray energy $E_{\rm X, iso}=10^{53} \rm erg$ is used.

2. The closure relation analysis suggests the CNM density profile of AT2022cmc as $n\,\propto \,R^{-1.47}$. The detailed study with FS model fit finds $n\,\propto \,R^{-1.68}$. The results are similar to that of Sw J1644+57 and are consistent with a Bondi-like accretion in history. The nucleus of AT2022cmc may be a low-luminosity AGN, but its luminosity is much weaker than the observed long-lasting optical ``plateau'' luminosity. 

3. Compared with the one-component jet predictions, the excess in the early sub-millimeter and radio data support a two-component jet model (a narrow-fast component and a wide-slow component). The best-fit parameter values of the two-component jet model are given in Table\,\ref{tab:fit-2comp}. Such a two-component jet structure is also revealed in the first jetted TDE candidate Sw J1644+57, indicating similar jet physics in these two events.

4. AT2022cmc provides a good prototype for unveiling jetted TDEs from optical and X-ray surveys. Such kind of events can be well monitored by EP/FXP even if it is viewed off-axis with view angle $\theta_{\rm obs} \leq 36.3^{\circ}$.

%$\theta_{\rm j}$.

%4. $n_{18}$

%5. $p$

%6. $\epsilon_{e}, \epsilon_{\rm B}$

%The free parameters of the afterglow are given in Table~\ref{tab:fit}, We find that the r band and the radio band data of AT2022cmc is well explained by the forward shock model as shown in Fig.~\ref{fig:model_lc}, 

\section*{acknowledgments}
We are very grateful to Yanan Wang, Lixin Dai, Hui Li, Bing Zhang, Yuan-Chuan Zou, He Gao, and Jumpei Takata for their helpful discussions. This work is supported by the National Key R\&D Program of China (Nos. 2020YFC2201400), and the National Natural Science Foundation of China under grants U2038107, U1931203, and 12021003. D.X. acknowledges support by the science research grants from the China Manned Space Project with NO. CMS-CSST-2021-A13 and CMS-CSST-2021-B11. W. H. Lei acknowledges support by the science research grants from the China Manned Space Project with NO.CMS-CSST-2021-B11. W. Xie acknowledges support by the Science and Technology Foundation of Guizhou Province (grant No. QianKeHeJiChu
ZK[2021]027). The authors acknowledge Beijing PARATERA Tech CO., Ltd. for providing HPC resources that have contributed to the research results reported within this paper.

\bibliography{ref}{}
\bibliographystyle{aasjournal}



\end{document}